\documentclass[paper]{geophysics}
\pdfoutput=1
\usepackage{amsmath}
\usepackage{amssymb}
\usepackage{amsfonts}
\usepackage{color}
\usepackage{array}
\usepackage{seg}

\begin{document}

\renewcommand{\figdir}{.} 

\title{Artifact reduction in pseudo-acoustic modeling by pseudo-source injection}
\author{Musa Maharramov}
\righthead{Pseudoacoustic modeling}
\lefthead{Maharramov}
\footer{SEP--152}
\maketitle

\begin{abstract}
I provide a framework for deriving fast finite-difference algorithms for the numerical modeling of acoustic wave propagation in anisotropic media. I deploy it in the case of transversely isotropic media to implement a kinematically accurate fast finite-difference modeling method. This results in a significant reduction of the shear artifacts compared to similar kinematically accurate finite-difference methods.
\end{abstract}

\section*{Introduction}

Transverse isotropy and orthorhombic media are of significant interest for industrial applications \cite[]{Grechka}.

The \emph{pseudo-acoustic} method of \cite[]{Alkhalifah98} is the anisotropic counterpart of isotropic acoustic modeling. However, this and similar anisotropic finite-difference methods suffer from \emph{shear artifacts} or are based on approximations that break down for strong anisotropy \cite[]{FDF10}, \cite[]{ZPS12}, (I note that both references discuss transverse isotropy but similar challenges exist for finite-difference modeling in  orthorhombic media). 

The objective of this work is to propose a computationally efficient finite-difference wave propagation modeling method for the vertically transversely isotropic (VTI) media that should be largely free of shear artifacts. Although I, too, demonstrate the method for VTI media, the concept extends to the orthorhombic case and the corresponding tilted symmetries.

Derivation of pseudo-acoustic (systems of) equations for a specific medium symmetry can be described as a three-step process:
\begin{enumerate}
        \item[1)] Derive a phase velocity surface \cite[]{Musgrave} as a function of the angle of propagation.
        \item[2)] Derive a dispersion relation from 1) \cite[]{Alkhalifah98}.
        \item[3)] Interpret the dispersion relation as an evolutionary pseudo-differential equation, and transform it into a form suitable for numerical solution.
\end{enumerate}
The cause of numerical artifacts is that the pressure and shear wave velocity surfaces remain coupled after deriving computationally feasible equations in step 3 (more specifically, the pressure mode and \emph{one} of the shear modes remain coupled).

My method can be summarized as follows:
\begin{enumerate}
        \item[2$^\prime$)] After step 1) above, extract the branch of the phase velocity surface corresponding to the pressure wave velocity.
        \item[3$^\prime$)] Approximate the resulting $V^2=F(\mathbf{m},\boldsymbol\theta)$, where $V$ is the pressure wave velocity, $\mathbf{m}$ stands for medium parameters, and $\boldsymbol\theta$ is the propagation direction, with a computationally efficient numerical Fourier operator. This can be, e.g., a \emph{trigonometric polynomial} in $\theta$ \cite[]{Iserles}, with coefficients depending on $\mathbf{m}$, as practiced in some of the existing spectral pseudo-acoustic modeling methods \cite[]{Etgen}, or a pseudo-differential operator spatially \emph{constrained to a narrow depth range} of sources and receivers, as demonstrated in this paper.
\item[4)] Derive a \emph{coupled} pseudo-pressure, pseudo-shear differential equation system analogous to, e.g., \cite{Alkhalifah2000}.
\item[5)] At each time step apply the spatial component of the pseudo-differential operator derived in step 3$^\prime$) to the injected source\footnote{or receiver data if back-propagating receivers for, e.g., reverse time migration} using a spectral method with spatial interpolation.
\item[6)] Inject the result of 5) as a ``pseudo-source'' into the second component of the system derived in 4), while injecting the true source into the primary component. 
\end{enumerate}
Step 6) assumes a VTI anisotropy, and that the system described in step 4) is that of \cite[]{Alkhalifah2000}. For equivalent alternative systems for VTI media \cite[]{FDF10}, or for other types of anisotropy, the injected sources in step 6) will be linear combinations of the true source and pseudo-source. 

\plot{model1}{width=.7\columnwidth}
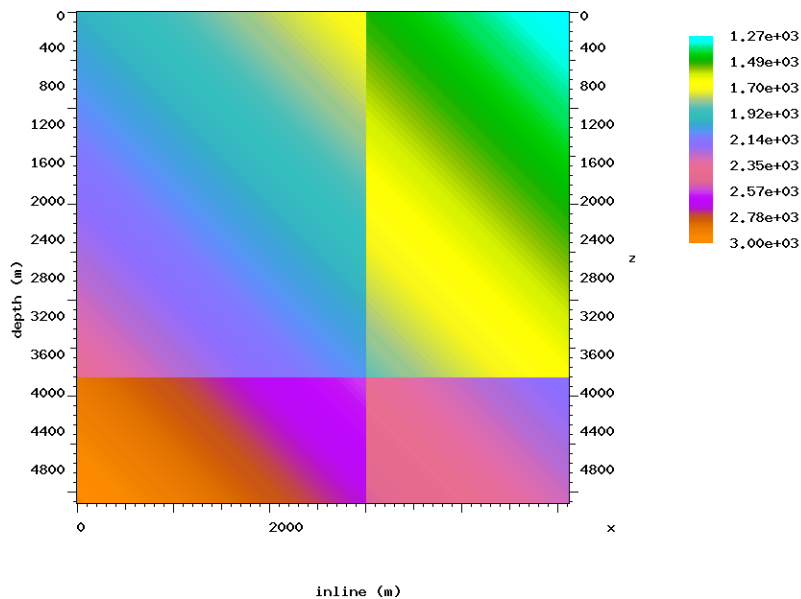
{Test model with smooth and sharp $V_P$ gradients and constant $\epsilon=0.3$ and $\delta=0.1$.}

\plot{model2}{width=.7\columnwidth}
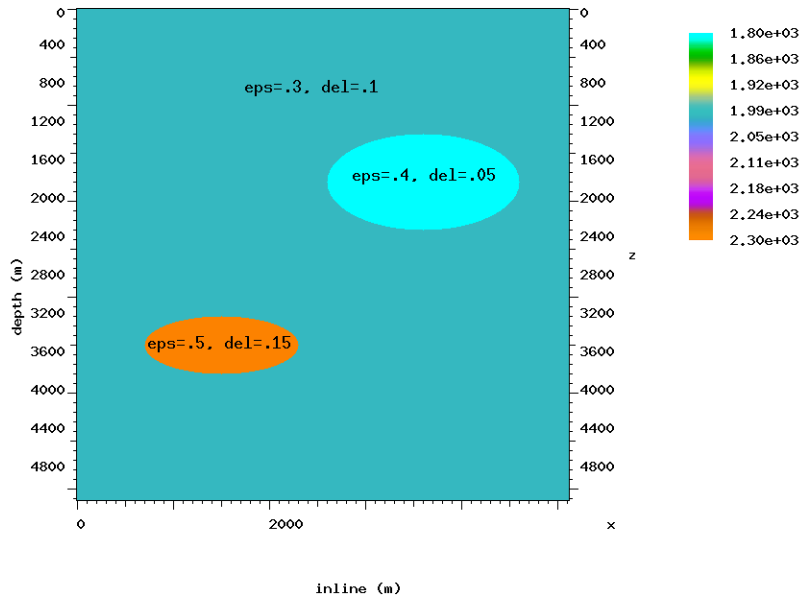
{Test model with two anisotropic inclusions.}

\section{The pseudo-differential modeling operator}
In step 1) we start with the equation for $V(\theta)$ in a VTI medium \cite[]{Tsvankin96}
        \begin{equation}
                \begin{aligned}
                        &               \frac{V^2(\theta)}{V_P^2}\;=\;1+\epsilon \sin^2\theta-\frac{f}{2}\pm
                        \frac{f}{2}\sqrt{\left(1+\frac{2\epsilon \sin^2\theta}{f}\right)^2-\frac{2(\epsilon-\delta)\sin^2 2\theta}{f}},\\
                &        \text{ with }f\; =1-\frac{V^2_S}{V^2_P},\text{ and}\\
                &        \sin\theta\;=\;\frac{V(\theta)\left[k_x\right]}{\left[\frac{\partial}{\partial t}\right]},\;\;\cos\theta\;=\;\frac{V(\theta)\left[k_z\right]}{\left[\frac{\partial}{\partial t}\right]}
        \end{aligned}
        \label{eq:v}
\end{equation}
where $V_P$ and $V_S$ are vertical pressure and shear wave velocities and $\epsilon$ and $\delta$ are the Thomsen parameters \cite[]{Thomsen86}. We assume that $V_S=0$, as we are not interested in propagating shear modes, thus $f=1$. Note that here and in the subsequent analysis we consider two-dimensional VTI, however, the results naturally extend to three dimensions by identifying $k_x$ with the \emph{radial} wavenumber---see, e.g., \cite[]{MUSAEAGE11}. I use the equivalence $k_u=-i\frac{\partial}{\partial u}$ in (\ref{eq:v}), where $u$ is an arbitrary variable, to stress that the phase velocity equation can be interpreted as both a dispersion relation and a pseudo-differential operator. In step 2$^\prime$), we extract the branch of the square root with the positive sign in (\ref{eq:v}), corresponding to the (higher) pressure wave velocity. The resulting dispersion relation can be interpreted as an evolutionary pseudo-differential operator governing \emph{kinematically accurate} propagation of the pressure wave:
 \begin{equation}
                \begin{aligned}
   & \frac{\partial^2}{\partial t^2} - V^2_P\left(\overset{2}{z},\overset{2}{x}\right)\frac{\Delta}{2}  - \epsilon\left(\overset{2}{z},\overset{2}{x}\right) V^2_P\left(\overset{2}{z},\overset{2}{x}\right)\frac{\partial^2 }{\partial x^2}  \;=\; \\
 &  V^2_P\left(\overset{2}{z},\overset{2}{x}\right) \frac{\Delta}{2}\sqrt{\left[1+ 2\epsilon\left(\overset{2}{z},\overset{2}{z}\right) \frac{\partial ^2}{\partial x^2} \frac{1}{\Delta}\right]^2-8\left(\epsilon\left(\overset{2}{z},\overset{2}{x}\right)-\delta\left(\overset{2}{z},\overset{2}{x}\right)\right) \frac{\partial ^2}{\partial x^2}  \frac{\partial ^2}{\partial z^2}\frac{1}{\Delta^2}} ,
        \end{aligned}
        \label{eq:pdo}
\end{equation}
where \[\Delta\;=\; \frac{\partial^2}{\partial x^2} +\frac{\partial^2}{\partial z^2} \] is the Laplace operator, and``2'' over $x$ and $z$ means that the multiplication by functions of spatial variables follows the application of differential operators in the pseudo-differential operator sense \cite[]{Maslov}. This is equivalent to ``freezing'' the operator coefficients, or assuming local homogeneity. Solving (\ref{eq:pdo}) for arbitrary heterogeneous media may be numerically challenging, because the Thomsen parameters $\epsilon(z,x)$ and $\delta(z,x)$ appear inside the square root of a pseudo-differential operator. However, operator (\ref{eq:pdo}) may simplify numerically if it is  applied to a function with spatially bounded support -- e.g., a source wavelet or receiver data. An alternative to solving the full pseudo-differential operator equation (\ref{eq:pdo}) is to approximate, in step 3$^\prime$), the extracted pressure velocity branch with a trigonometric polynomial:
\begin{equation}
        V^2(\theta)\;\approx\;V^2_P\sum_{n=0}^{N}{a_n\sin^{2n}(\theta)},
        \label{eq:approx}
\end{equation}
where the coefficients $a_n,\;n=0,\ldots,N$ depend on medium parameters. From the last line of (\ref{eq:v}) we can see that velocity surface (\ref{eq:approx}) translates into the following pseudo-differential operator equation
\begin{equation}
        \frac{\partial^2}{\partial t^2}\;=\;V^2_P\sum_{n=0}^{N}{a_n \frac{\partial^{2n}}{\partial x^{2n}}\Delta^{1-n}},
        \label{eq:pdo2}
\end{equation}
Equation (\ref{eq:pdo2}) can be solved by applying the operators \[\frac{\partial^{2n}}{\partial x^{2n}}\Delta^{1-n} \] to the wave field in the spatial Fourier domain, then summing up the results with spatially-dependent coefficients $a_n$ in the spatial domain. Important particular cases of approximation (\ref{eq:approx}) are the \emph{weak anisotropy} approximation \cite[]{Grechka}
\begin{equation}
        V^2(\theta)\;\approx\;V^2_P\left(1+\delta\sin^2\theta + \frac{\epsilon-\delta}{1+2\delta}\sin^4\theta\right),
                \label{eq:waa}
\end{equation}
and the VTI approximation due to Harlan and Lazear \cite[]{Harlan98} used by \cite{Etgen}
        \begin{equation}
                V^2(\theta)\;=\;V^2_P\cos^2\theta + \left(V^2_{P\text{NMO}}-V^2_{P\text{Hor}}\right)\cos^2\theta\sin^2\theta+V^2_{P\text{Hor}}\sin^2\theta,
                \label{eq:Harlan}
        \end{equation}
        where the subscripts $P\text{Hor}$ and $P\text{NMO}$ denote the horizontal and NMO pressure wave velocities, respectively. Note that both (\ref{eq:waa}) and (\ref{eq:Harlan}) correspond to $N=2$ in (\ref{eq:approx}) and are suitable for weakly anisotropic VTI but break down in strong anisotropy. The case of $N=3$ requires one additional inverse FFT for VTI but is accurate for a wide range of Thomsen parameters within (and beyond) practical requirements. Adapting (\ref{eq:approx}) for TTI media would require the application at each time step of 5 additional inverse FFTs for $N=2$ and extra 16 inverse FFTs for $N=3$. 
        
        Solving (\ref{eq:pdo2}) for $N=2,3$ using the described spectral method is an efficient modeling method in its own right, especially for VTI media where the number of FFTs at each time step is very low. However, in the next section I describe a finite-difference method that can outperform the spectral method for complex media and conceptually generalizes for other kinds of anisotropy.

\plot{p1fdnored}{width=.7\columnwidth}
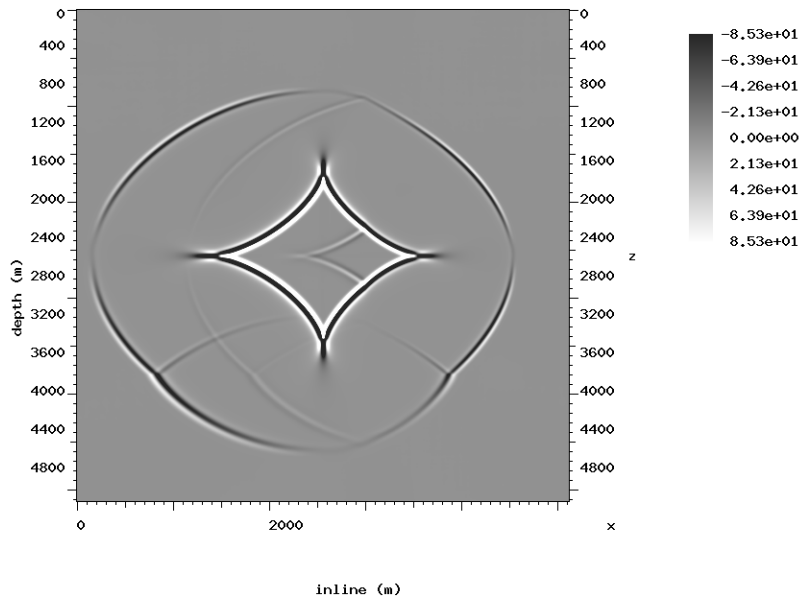
{Shear artifacts in the solution of (\ref{eq:coup}) for the model of Figure~\ref{fig:model1} with sources injected in component $r$.}

\plot{q1fdpdoqsource}{width=.7\columnwidth}
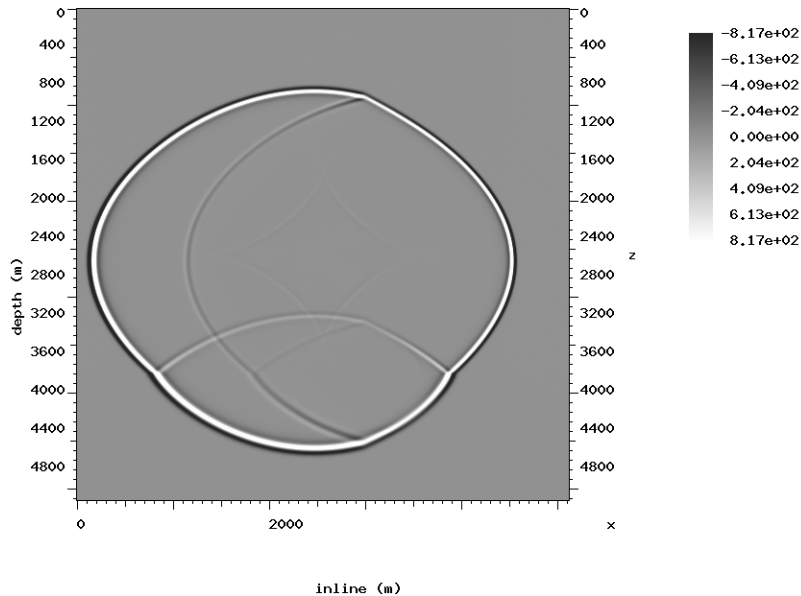
{Shear artifacts in the solution of  (\ref{eq:coup}) with sources injected in component $q$.}

\section{The Finite-Difference Method}

In step 4) we square the pseudo-differential operator equation (\ref{eq:pdo}) so as to get rid of the square root, and obtain the following system of coupled second-order partial differential equations \cite[]{Alkhalifah2000}:
\begin{equation}
\begin{aligned}
& \frac{\partial^2 q}{\partial t^2}\;=\;V^2_{P\text{Hor}}\frac{\partial^2 q}{\partial x^2}+V^2_P\frac{\partial^2 q}{\partial z^2} +V^2_P\left(V^2_{P\text{Hor}}-V^2_{P\text{NMO}}\right)\frac{\partial^4 r}{\partial x^2 \partial z^2},\\
& \frac{\partial^2 r}{\partial t^2}\;=\;q,
\end{aligned}
\label{eq:coup}
\end{equation}
where $r(z,x,t)$ and $q(z,x,t)$ are the pressure field and its second temporal derivative, and \[ V_{P\text{Hor}}(z,x)\;=\;V_P(z,x)\sqrt{1+2\epsilon(z,x)},\; V_{P\text{NMO}}(z,x)\;=\;V_P(z,x)\sqrt{1+2\delta(z,x)}. \]
Since the resulting system now includes the branch with the negative square root in (\ref{eq:v}), solution of this system \emph{may} suffer from shear artifacts as shown in Figure~\ref{fig:p1fdnored}. The artifacts can be reduced by injecting sources in to the second component $q$ \cite[]{FDF10}; however, they are still present---see Figure~\ref{fig:q1fdpdoqsource}. However, the pseudo-differential operator equation (\ref{eq:pdo}) can be used to reduce the unwanted artifacts (appearing as the ``diamond''-shaped inverted wavefront in the figure). Equation (\ref{eq:v}) and the corresponding pseudo-differential equation do not describe any pressure to shear conversion but rather govern the \emph{independent} propagation of the pressure and shear waves. The same is true of the ``coupled'' system of differential equations. Consequently, any shear artifacts that appear in a solution to the coupled system of differential equations is likely due to the \emph{pseudo-shear modes present in the wave field}. We can use the fact that the system of two coupled equations requires injecting \emph{two} sources, to \emph{manufacture} a \emph{pseudo-source} to be injected into one of the components so as to suppress the shear modes. More specifically, if $\phi(z,x,t)$ is a time-dependent source function, then at each time step component $r$ is injected with $\phi$, and component $q$ is injected with the result of applying the \emph{spatial part} of the pseudo-differential operator (\ref{eq:pdo}) to $\phi(z,x,t)$:
 \begin{equation}
                \begin{aligned}
   & r(z,x,t_n) \;=\; r(z,x,t_n) + \phi(z,x,t_n),\\
   & q(z,x,t_n) \;=\; q(z,x,t_n) +  V^2_P\left\{  \left(\overset{2}{z},\overset{2}{x}\right)\frac{\Delta}{2}  + \epsilon\left(\overset{2}{z},\overset{2}{x}\right) \frac{\partial^2 }{\partial x^2}  + \right.\\
 & \left. + \frac{\Delta}{2}\sqrt{\left[1+ 2\epsilon\left(\overset{2}{z},\overset{2}{z}\right) \frac{\partial ^2}{\partial x^2} \frac{1}{\Delta}\right]^2-8\left(\epsilon\left(\overset{2}{z},\overset{2}{x}\right)-\delta\left(\overset{2}{z},\overset{2}{x}\right)\right) \frac{\partial ^2}{\partial x^2}  \frac{\partial ^2}{\partial z^2}\frac{1}{\Delta^2}} \right\} \phi,
        \end{aligned}
        \label{eq:pdosource}
\end{equation}
followed by a finite-difference time propagation step of system (\ref{eq:coup}). This procedure ensures that the two-component source in the right-hand side of (\ref{eq:pdosource}) satisfies equation (\ref{eq:pdo}). Since solutions of (\ref{eq:pdo}) are shear-free, the injected sources will not give rise to shear modes because the solution of (\ref{eq:coup}) is effectively projected on to the space of solutions of (\ref{eq:pdo}). 

\section{Numerical Examples}

Figure~\ref{fig:p1fdpdo} shows the result of applying the pseudo-source finite-difference method to the propagation in a heterogeneous VTI medium described by the model of Figure~\ref{fig:model1}, with a Ricker source. The corresponding result obtained by solving the full pseudo-differential operator equation (\ref{eq:pdo}) is shown in Figure~\ref{fig:p1pdo}. Note the significant reduction of the shear artifacts, and that although we use the full pseudo-differential operator for generating the pseudo-source in (\ref{eq:pdosource}), the fact that the source is localized makes this computationally efficient, obviating the need for approximations like (\ref{eq:pdo2}).

\plot{p1fdpdo}{width=.7\columnwidth}
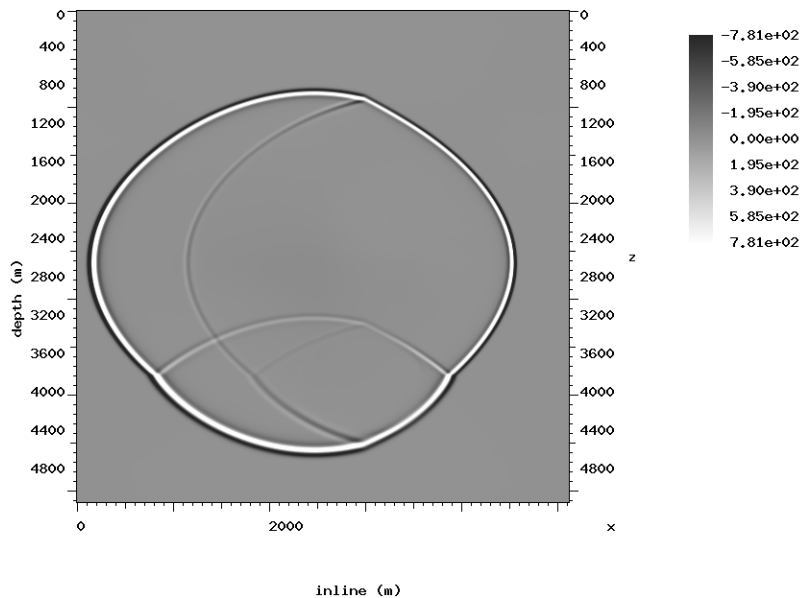
{Solution of (\ref{eq:coup}) for the model of Figure~\ref{fig:model1} with shear-reducing pseudo-sources. Note the good agreement with the result of solving the full pseudo-differential operator equation (\ref{eq:pdo}) in Figure~\ref{fig:p1pdo}.}

\plot{p1pdo}{width=.7\columnwidth}
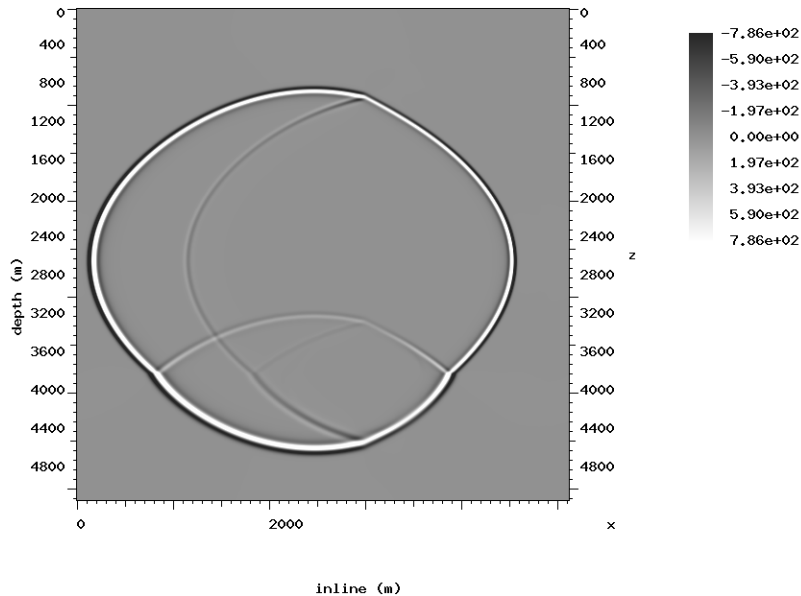
{Solution of the full pseudo-differential operator equation (\ref{eq:pdo}) for the model of Figure~\ref{fig:model1}. Note the good agreement with the result of Figure~\ref{fig:p1fdpdo}.}

The model of Figure~\ref{fig:model1}, while featuring both sharp and smooth vertical velocity variation, assumes constant $\epsilon=0.3$ and $\delta=0.1$. While adding the pseudo-source (\ref{eq:pdosource}) ensures that the solution of the coupled system (\ref{eq:coup}) stays within the space of solutions of (\ref{eq:pdo}) \emph{in the continuous limit} $\Delta t\to 0$, sharp contrasts in $\epsilon$ and $\delta$ may introduce numerical approximation errors that may contain a non-negligible shear component. Indeed, applying the method to the model of Figure~\ref{fig:model2}, featuring two inclusions with significantly different Thomsen parameters, we can see weak artifacts (single lines) within the inclusions in Figure~\ref{fig:markedp3fdpdo} for the finite-difference method that are absent from the result in Figure~\ref{fig:p2pdo} obtained by solving the full pseudo-differential operator (\ref{eq:pdo}). Figure~\ref{fig:p2fdpdo} shows the result of using the finite-difference method with pseudo-sources after smoothing the $\epsilon$ and $\delta$ models. Note the vertical velocity model $V_P$ was not smoothed. The result shows that the artifacts within the inclusions were almost completely removed.

\vspace{-.5cm}
\plot{markedp3fdpdo}{width=.9\columnwidth}
{
\vspace{-.05cm}
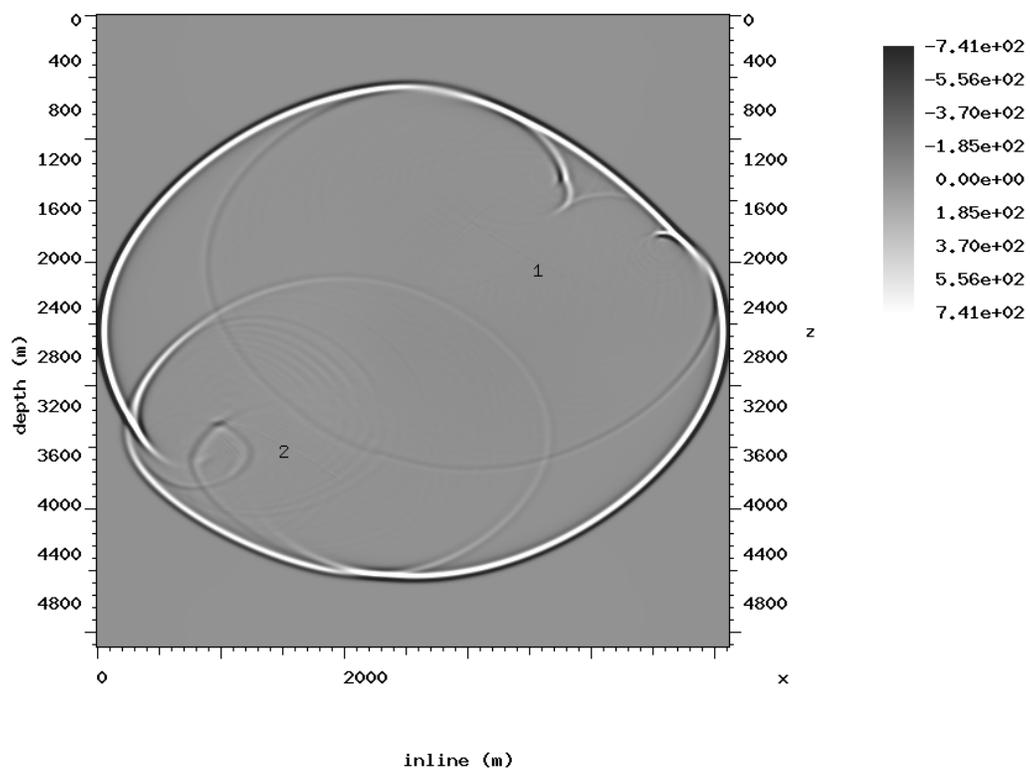
Solution of (\ref{eq:coup}) for the model of Figure~\ref{fig:model2} with shear-reducing pseudo-sources. A sharp contrast in the values of Thomsen parameters at the inclusion boundaries results in weak artifacts (single lines) within the inclusions. }

\plot{p2pdo}{width=.7\columnwidth}
{Solution of the full pseudo-differential operator equation (\ref{eq:pdo}) for the model of Figure~\ref{fig:model2}. }

\plot{p2fdpdo}{width=.7\columnwidth}
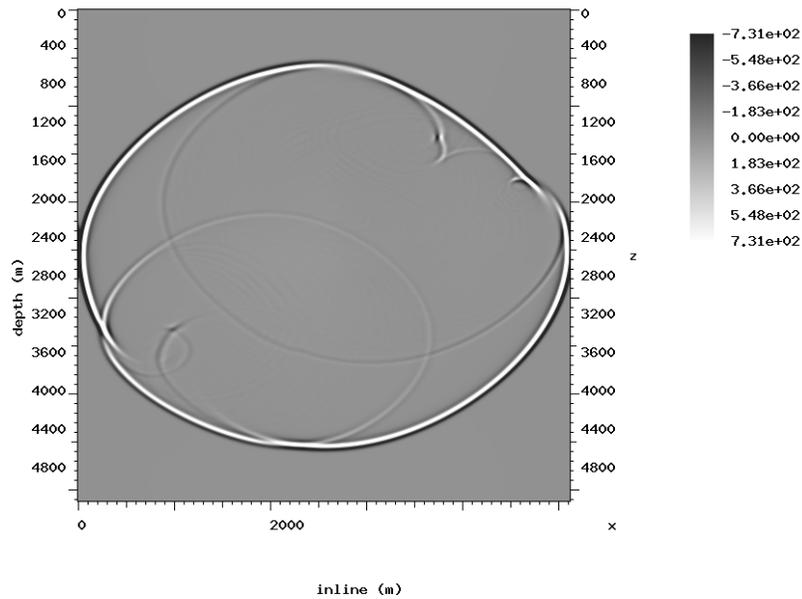
{Solution of (\ref{eq:coup}) for the model of Figure~\ref{fig:model2} with shear-reducing pseudo-sources. Smoothing of the Thomsen parameters resulted in weaker artifacts within the inclusions (compare with Figure~\ref{fig:markedp3fdpdo}). No smoothing was applied to $V_P$. }

\section{Conclusions and Perspectives}

The proposed pseudo-source finite-difference method allows us to take advantage of the computationally cheap finite-difference solvers for the traditional pseudo-acoustic (fourth-order) systems while achieving a significant reduction of shear artifacts. The method is kinematically accurate for VTI media, and can be extended in principle to other kinds of anisotropy. While my implementation is based on using the coupled system (\ref{eq:coup}) of \cite{Alkhalifah2000}, the method can be adapted to use equivalent systems \cite[]{FDF10}. In that case the two-component source becomes a linear combination of the true source and the pseudo-source terms, with the coefficients of the linear combination determined by the relationship between the solution of the equivalent system and that of system (\ref{eq:coup}).

\section{Acknowledgements}

I would like to thank Stewart Levin for a number of useful discussions.

\bibliographystyle{seg}
\bibliography{musavti}

\end{document}